\documentclass[11pt]{article}

\usepackage{hyperref}
\usepackage[a4paper,text={6in,9in},centering]{geometry} 
\usepackage{amsmath}
\usepackage{amssymb}
\usepackage{amsbsy}
\usepackage{mathrsfs}
\usepackage{graphicx}
\usepackage{subfigure}
\usepackage{epic,eepic,xcolor} 	
\usepackage{cite}
\makeatletter

\@addtoreset{equation}{section}
\makeatother

\title{
	\vskip -3em
	\begin{flushright}
	\small
	IFUP-TH/2011-9
	\end{flushright}
	\vskip 5em
	\textbf{\large Path Integral on Star Graph}
}
\author{
Satoshi Ohya$^{\,a),\,b),}$\footnote{Present address: Harish-Chandra Research Institute, Chhatnag Road, Jhusi, Allahabad 211 019, India.
\texttt{E-mail:\href{mailto:ohya@hri.res.in}{ohya@hri.res.in}}}\\[2ex]
\textit{\small $^{a)}$Istituto Nazionale di Fisica Nucleare -- Sezione di Pisa}\\
\textit{\small Largo Bruno Pontecorvo 3, 56127 Pisa, Italy}\\
\textit{\small $^{b)}$Dipartimento di Fisica ``Enrico Fermi'', Universit\`a di Pisa}\\
\textit{\small Largo Bruno Pontecorvo 3, 56127 Pisa, Italy}\\[1ex]
\texttt{\small E-mail:\href{mailto:satoshi.ohya@pi.infn.it}{satoshi.ohya@pi.infn.it}}
}
\date{\small (Dated: \today)}

\begin{document}
\maketitle
\thispagestyle{empty}
\begin{abstract}
In this paper we study path integral for a single spinless particle on a star graph with $N$ edges, whose vertex is known to be described by $U(N)$ family of boundary conditions.
After carefully studying the free particle case, both at the critical and off-critical levels, we propose a new path integral formulation that correctly captures all the scale-invariant subfamily of boundary conditions realized at fixed points of boundary renormalization group flow.
Our proposal is based on the folding trick, which maps a scalar-valued wave function on star graph to an $N$-component vector-valued wave function on half-line.
All the parameters of scale-invariant subfamily of boundary conditions are encoded into the momentum independent weight factors, which appear to be associated with the two distinct path classes on half-line that form the cyclic group $\mathbb{Z}_{2}$.
We show that, when bulk interactions are edge-independent, these weight factors are generally given by an $N$-dimensional unitary representation of $\mathbb{Z}_{2}$.
Generalization to momentum dependent weight factors and applications to worldline formalism are briefly discussed.
\end{abstract}

\newpage
\section{Introduction} \label{sec:1}
Over the last one and a half decade, quantum graphs have been attracted much attention as the simplest low-energy effective theory for mesoscopic networks of one-dimensional quantum wires, whose configuration space is effectively modeled by a graph (see for a brief survey \cite{Kuchment:2008} and references therein).
In general, a quantum graph is a non-differentiable manifold and has singularities at junction points (vertices) of graph edges.
A standard lore for taming such singularity is first to require Kirchhoff's law of conserved currents (Noether currents) at the vertex and then to reduce to the problem of boundary conditions for wave functions/quantum fields defined in the bulk.

Although Kirchhoff's law of conserved currents has been vastly studied in the literature in the operator formalism, its path integral description is not yet fully understood.
The problem is that boundary conditions provided by Kirchhoff's law of conserved currents will be characterized by a large number of parameters, however, there is no \textit{a priori} prescription to include these parameters into the path integral weight $\exp(iS)$ nor into the path integral measure $\mathcal{D}x$.
Although there exist several attempts, most of the literature concern path integral description for a system defined on a half-line with the Robin boundary condition \cite{Clark:1980xt,Farhi:1989jz,Gamboa:1995ew,Bastianelli:2006hq,Bastianelli:2007jr,Asorey:2007zza,Bastianelli:2008vh}, which is characterized by only one real parameter, or a system on a line with a Dirac $\delta$-potential \cite{Goovaerts:1973,Bauch:1985,Gaveau:1986,Lawande:1988,Blinder:1988,Grosche:1990um,Vinas:2010ix}, which is characterized by a single real coupling constant, or a system on a circle with a single scale-independent point interaction \cite{Fulop:1999pf,Fulop:2003,Asorey:2007zza}, which is characterized by two real parameters, or a system on a circle with a single $U(2)$ family of point interactions \cite{Carreau:1990wh,Ohya:2009yg}, which is characterized by four real parameters.
Path integral approach to more complicated quantum graph that contains more boundary condition parameters is yet mysterious, even for the case where there is no interaction in the bulk.
In this paper we would like to tackle with this problem.
To get the feel of our problem, in the first part of this paper we will concentrate ourselves to non-relativistic quantum mechanics for a free spinless particle on a star graph with $N$ edges, whose vertex is known to be described by $U(N)$ family of boundary conditions and hence characterized by $N^{2}$ real parameters.
Since the Schr\"odinger equation on star graph is a well-defined problem, the most unambiguous way to find the path integral formulation on star graph will be as follows: (i) first solve the Schr\"odinger equation with $U(N)$ family of boundary conditions, (ii) and then find the complete orthonormal set of energy eigenfunctions, (iii) and then evaluate the Feynman kernel in terms of the complete orthonormal set, (iv) and finally rewrite the kernel into the path integral representation.
With this approach we will show that all the $N^{2}$ parameters are naturally encoded into the weight factors of scattering matrix (S-matrix) associated with the two distinct path classes that form the cyclic group of order $2$ ($\mathbb{Z}_{2}$).
In the second part of this paper we will proceed to include edge-independent bulk interactions and show that, when the theory lies on a fixed point of boundary renormalization group (RG) flow, the weight factors are generally given by an $N$-dimensional unitary representation of $\mathbb{Z}_{2}$.

The rest of this paper is organized as follows.
In Section \ref{sec:2} we first recall the complete orthonormal set of energy eigenfunctions of the free Hamiltonian on star graph and then compute the Feynman kernel.
By taking the continuum limit of time-slicing representation of the kernel, we find that the path integral description of $U(N)$ family of boundary conditions is naturally formulated into the phase space path integral rather than configuration space path integral with exponential weighting factor $\exp(iS_{\text{free}}[x(t), p(t)])$, where $S_{\text{free}}$ is the action for a free particle in the Hamiltonian formulation, and another weighting factor of S-matrix $\mathbb{S}(p)$ that depends on the initial momentum $p = p(0)$.
The resultant path integral becomes a summation over two distinct path classes categorized by the number of interactions ($n=0$ or $1$) for which a classical particle experiences at the vertex.
Contributions from the path class of $n=1$ turn out to be weighted by the S-matrix.
In Section \ref{sec:3} we discuss the critical case; that is, the case of boundary conditions realized at the fixed points of boundary RG flow.
In this critical case the S-matrix becomes constant matrix such that the phase space path integral can be simply reduced to the ordinary configuration space path integral.
In Section \ref{sec:4} we generalize to an interacting case by adding an edge-independent bulk interaction.
We show that, at the fixed point of boundary RG flow, the weight factors are generally given by an $N$-dimensional unitary representation of the cyclic group $\mathbb{Z}_{2}$, which is the main result of this paper.
Section \ref{sec:5} is devoted to conclusions, discussions for remaining issues and a simple application to worldline formalism on star graph.
Computational details for the proof of orthonormality and completeness of the eigenfunctions of the free Hamiltonian on star graph are presented in Appendix \ref{appendix:A}.

\section{Free particle} \label{sec:2}
\begin{figure}[t]
\centering
\subfigure[Unfolding picture.]{\includegraphics{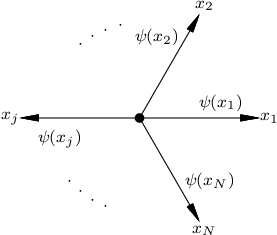} \label{fig:1a}}
\raisebox{1.9cm}{$\quad\xrightarrow[]{\text{folding}}\quad$}
\subfigure[Folding picture.]{\raisebox{1.7cm}{\includegraphics{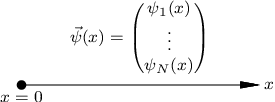}} \label{fig:1b}}
\caption{Star graph with $N$ edges. In the folding picture wave function of the system is given by an $N$-component vector of $L^{2}(\mathbb{R}^{+})$-functions.}
\label{fig:1}
\end{figure}

To begin with, let us first recall non-relativistic quantum mechanics for a free particle on a star graph with $N$ edges.
The star graph $\Gamma$ is the one-point union of $N$ positive half-lines $\mathbb{R}^{+}_{i} = \{x_{i} \in \mathbb{R} \mid x_{i} > 0\}$ ($i=1,\cdots,N$) such that the Hilbert space $\mathcal{H}(\Gamma)$ is just given by the direct sum $\mathcal{H} = \bigoplus_{i=1}^{N}L^{2}(\mathbb{R}^{+}_{i})$.
Since $L^{2}(\mathbb{R}^{+}_{i})$ is equivalent to $L^{2}(\mathbb{R}^{+}_{j})$ for any $i, j$, the Hilbert space can also be written as follows:
\begin{align}
\mathcal{H}
&= L^{2}(\mathbb{R}^{+}) \otimes \mathbb{C}^{N}. \label{eq:2.1}
\end{align}
It should be emphasized that the concept behind the change of notation from $\bigoplus_{i=1}^{N}L^{2}(\mathbb{R}^{+}_{i})$ to $L^{2}(\mathbb{R}^{+}) \otimes \mathbb{C}^{N}$ is the so-called folding trick \cite{Wong:1994pa,Oshikawa:1996,Bachas:2001vj,Bajnok:2004jd}, which maps a single ``scalar-valued'' wave function on star graph to an $N$-component vector-valued wave function on half-line; see Figure \ref{fig:1}.
In the folding picture the wave function (element of $\mathcal{H}$) is then given by
\begin{align}
\Vec{\psi}(x)
&= 	\bigl(\psi_{1}(x), \cdots, \psi_{N}(x)\bigr)^{T}, \label{eq:2.2}
\end{align}
where $\psi_{j}(x)$ ($j=1,\cdots,N$) corresponds to the wave function on the $j$th edge with $x$ being the distance from the vertex.
$T$ stands for transposition.
The time-independent Schr\"odinger equation for a free particle now becomes the following matrix equation
\begin{align}
\mathbb{H}_{\text{free}}\Vec{\psi}(x)
&= 	E\Vec{\psi}(x), \quad
x \neq 0, \label{eq:2.3}
\end{align}
where the bulk Hamiltonian is given by the $N \times N$ diagonal matrix
\begin{align}
\mathbb{H}_{\text{free}}
&:= 	\mathrm{diag}(H_{\text{free}}, \cdots, H_{\text{free}})
= 	H_{\text{free}} \otimes \mathbb{I}, \quad
H_{\text{free}}
:= 	-\frac{\mathrm{d}^{2}}{\mathrm{d}x^{2}}, \label{eq:2.4}
\end{align}
with $\mathbb{I}$ being the $N \times N$ unit matrix.
In this paper we will work in the units where $\hbar = 2m = 1$.
As is well-known, Kirchhoff's law of probability currents (or self-adjoint extension of $\mathbb{H}_{\text{free}}$) leads to the following $U(N)$ family of boundary conditions at the vertex:
\begin{align}
(\mathbb{I} - \mathbb{U})\Vec{\psi}(0)
- iL_{0}(\mathbb{I} + \mathbb{U})\Vec{\psi}^{\prime}(0)
&= 	\Vec{0}, \quad
\mathbb{U} \in U(N), \label{eq:2.5}
\end{align}
where prime (${}^{\prime}$) indicates the derivative with respect to $x$.
$L_{0}$ is an arbitrary length parameter (or renormalization scale \cite{Ohya:2010zm}) which is inevitably introduced to adjust the length dimension of the equation \eqref{eq:2.5}.
In this paper we will use black board symbols ($\mathbb{A}$, $\mathbb{B}$, $\mathbb{C}$, etc.) to denote matrices.
For the following discussions it is convenient to parameterize the matrix $\mathbb{U} \in U(N)$ into the following spectral decomposition form
\begin{align}
\mathbb{U}
&= 	\sum_{a = 1}^{N}\mathrm{e}^{i\alpha^{a}}\mathbb{P}^{a}, \quad
\mathbb{P}^{a}
:= 	\Vec{\xi}^{a} \cdot (\Vec{\xi}^{a})^{\dagger}, \label{eq:2.6}
\end{align}
where $\mathrm{e}^{i\alpha^{a}}$ ($0 \leq \alpha^{a} < 2\pi$) is the $a$th eigenvalue of $\mathbb{U}$ and $\Vec{\xi}^{a} = ({\xi^{a}}_{1}, \cdots, {\xi^{a}}_{N})^{T}$ is the corresponding $a$th eigenvector that fulfills the eigenvalue equation $\mathbb{U}\Vec{\xi}^{a} = \mathrm{e}^{i\alpha^{a}}\Vec{\xi}^{a}$, the orthonormality $(\Vec{\xi}^{a})^{\dagger} \cdot \Vec{\xi}^{b} = \delta^{ab}$ and the completeness $\sum_{a=1}^{N}\Vec{\xi}^{a} \cdot (\Vec{\xi}^{a})^{\dagger} = \mathbb{I}$.
With these eigenvectors $\mathbb{P}^{a}$ becomes the hermitian projection operator satisfying $\sum_{a=1}^{N}\mathbb{P}^{a} = \mathbb{I}$, $\mathbb{P}^{a}\mathbb{P}^{b} = \delta^{ab}\mathbb{P}^{b}$ and $(\mathbb{P}^{a})^{\dagger} = \mathbb{P}^{a}$, where dagger (${}^{\dagger}$) stands for the hermitian conjugate.
By substituting \eqref{eq:2.6} into the equation \eqref{eq:2.5}, the boundary condition is diagonalized and boils down to the following $N$ independent equations
\begin{align}
(\Vec{\xi}^{a})^{\dagger} \cdot \left[\Vec{\psi}(0) + L^{a}\Vec{\psi}^{\prime}(0)\right]
&= 	0, \quad
L^{a}
:= 	L_{0}\cot\frac{\alpha^{a}}{2}. \label{eq:2.7}
\end{align}

Now it is easy to find the solution to the Schr\"odinger equation \eqref{eq:2.3} with the boundary condition \eqref{eq:2.7}.
Let us first consider positive energy ($E>0$) solutions.
For given momentum $p=\sqrt{E}>0$, the continuum spectrum exhibits $N$-fold degeneracy and there exist $N$ distinct eigenfunctions which are orthogonal to each other yet have the same energy eigenvalue $E = p^{2}$.
One of the most convenient choices of such solutions is as follows:
\begin{align}
\Vec{\psi}^{a}(x; p)
&= 	\bigl[\mathbb{I}\exp(-ipx) + \mathbb{S}(p)\exp(+ipx)\bigr]\Vec{\xi}^{a} \nonumber\\
&= 	\left[
	\exp(-ipx) + \frac{ipL^{a} - 1}{ipL^{a} + 1}\exp(+ipx)
	\right]
	\Vec{\xi}^{a}, \quad 0<p<\infty, \quad a=1,\cdots, N, \label{eq:2.8}
\end{align}
where $\mathbb{S}(p)$ is the one-particle S-matrix given by
\begin{align}
\mathbb{S}(p)
&= 	\sum_{a=1}^{N}\frac{ipL^{a} - 1}{ipL^{a} + 1}\mathbb{P}^{a}
= 	\exp\left[i\sum_{a=1}^{N}2\mathrm{arccot}(pL^{a})\mathbb{P}^{a}\right], \label{eq:2.9}
\end{align}
whose $ii$-component gives the reflection coefficient for a particle residing on the $i$th edge and $ij$-component ($i \neq j$) the transmission coefficient for a particle traveling from the $j$th edge to the $i$th edge.
Note that the S-matrix satisfies the unitarity, $\mathbb{S}^{\dagger}(p)\mathbb{S}(p) = \mathbb{S}(p)\mathbb{S}^{\dagger}(p) = \mathbb{I}$, and the hermitian analyticity, $\mathbb{S}^{\dagger}(p) = \mathbb{S}(-p)$.
Consequently, $\Vec{\psi}^{a}(x; p)$ and $\Vec{\psi}^{a}(x; -p)$ are just related by the unitary transformation $\Vec{\psi}^{a}(x; p) = \mathbb{S}(p)\Vec{\psi}^{a}(x; -p)$.
Note also that any solution to the Schr\"odinger equation \eqref{eq:2.3} with the energy eigenvalue $E = p^{2}$ is given by the linear combination of $\Vec{\psi}^{a}(x; p)$, $a=1,\cdots,N$.

Let us next consider negative energy ($E < 0$) solutions, or bound states, which are characterized by the S-matrix pole lying on the positive imaginary $p$-axis.
As discussed in \cite{Ohya:2010zm}, there is a one-to-one correspondence between the eigenvalues of $\mathbb{U}$ and the negative energy spectrum: If the eigenphase $\alpha^{a}$ lies on the range $0<\alpha^{a}<\pi$ (or $0 < L^{a} < \infty$), there exists a normalizable bound state with the energy $E = -1/(L^{a})^{2}$.
If the eigenphase $\alpha^{a}$ lies on the range $\pi<\alpha^{a}<2\pi$ (or $-\infty < L^{a} < 0$), on the other hand, there exists a non-normalizable antibound state with the energy $E = -1/(L^{a})^{2}$.
The normalizable bound state solution that corresponds to the eigenvalue $\mathrm{e}^{i\alpha^{a}}$ is found to be of the form
\begin{align}
\Vec{\psi}_{B}^{a}(x)
&= 	\sqrt{\frac{2}{L^{a}}}\exp\left(-\frac{x}{L^{a}}\right)\Vec{\xi}^{a},
	\quad 0 < L^{a} < \infty. \label{eq:2.10}
\end{align}
The set of these eigenfunctions \eqref{eq:2.8} and \eqref{eq:2.10} provides the complete orthonormal basis of the system; that is, they satisfy the orthonormality
\begin{subequations}
\begin{align}
& 	\int_{0}^{\infty}\!\!\!\mathrm{d}x\,
	\bigl[\Vec{\psi}^{a}(x; p)\bigr]^{\dagger} \cdot \Vec{\psi}^{b}(x; q)
= 	2\pi\delta^{ab}\delta(p - q), \label{eq:2.11a}\\
& 	\int_{0}^{\infty}\!\!\!\mathrm{d}x\,
\bigl[\Vec{\psi}_{B}^{a}(x)\bigr]^{\dagger} \cdot \Vec{\psi}_{B}^{b}(x)
= 	\delta^{ab}, \label{eq:2.11b}\\
& 	\int_{0}^{\infty}\!\!\!\mathrm{d}x\,
	\bigl[\Vec{\psi}^{a}(x; p)\bigr]^{\dagger} \cdot \Vec{\psi}_{B}^{b}(x)
= 	\int_{0}^{\infty}\!\!\!\mathrm{d}x\,
\bigl[\Vec{\psi}_{B}^{a}(x)\bigr]^{\dagger} \cdot \Vec{\psi}^{b}(x; p)
= 	0, \label{eq:2.11c}
\end{align}
\end{subequations}
and the completeness
\begin{align}
\sum_{a=1}^{N}\int_{0}^{\infty}\!\frac{\mathrm{d}p}{2\pi}\,
\Vec{\psi}^{a}(x; p) \cdot \bigl[\Vec{\psi}^{a}(y; p)\bigr]^{\dagger}
+ \sum_{0<L^{a}<\infty}\Vec{\psi}_{B}^{a}(x) \cdot \bigl[\Vec{\psi}_{B}^{a}(y)\bigr]^{\dagger}
&= 	\delta(x - y)\mathbb{I}. \label{eq:2.12}
\end{align}
(For the computational details, see Appendix \ref{appendix:A}.)
\begin{figure}[t]
\centerline{\includegraphics{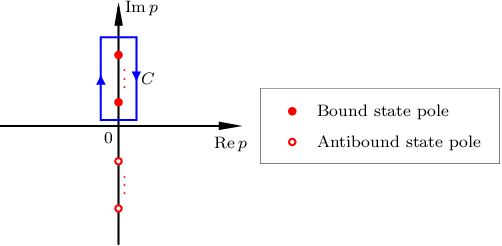}}
\caption{Integration contour $C$ and simple poles of one-particle S-matrix $\mathbb{S}(p)$ on the complex $p$-plain.}
\label{fig:2}
\end{figure} 
By using this complete orthonormal basis the Feynman kernel $\mathbb{K}(x, y; T) = \langle x|\mathrm{e}^{-i\mathbb{H}_{\text{free}}T}|y\rangle$ can be expressed as the following $N \times N$ matrix
\begin{align}
\mathbb{K}(x, y; T)
&= 	\sum_{a=1}^{N}\int_{0}^{\infty}\!\frac{\mathrm{d}p}{2\pi}\,
	\exp(-ip^{2}T)\Vec{\psi}^{a}(x; p) \cdot \bigl[\Vec{\psi}^{a}(y; p)\bigr]^{\dagger} \nonumber\\
&	+ \sum_{0<L^{a}<\infty}\exp\left[i\left(\frac{1}{L^{a}}\right)^{2}T\right]\Vec{\psi}_{B}^{a}(x) \cdot \bigl[\Vec{\psi}_{B}^{a}(y)\bigr]^{\dagger}, \label{eq:2.13}
\end{align}
whose $ij$-component gives the transition amplitude for a particle traveling from $(y, j)$ to $(x, i)$ in time interval $T$, where the notation $(x, i)$ stands for the position located at the distance $x$ from the vertex on the $i$th edge.
As easily checked the Feynman kernel \eqref{eq:2.13} satisfies the following equations:
\begin{subequations}
\begin{alignat}{3}
\text{(i)}&~
\text{Schr\"odinger equation}&\quad
&\bigl[i\partial_{T}\mathbb{I} - \mathbb{H}_\text{free}(-i\partial_{x})\bigr]\mathbb{K}(x, y; T) = 0;& \label{eq:2.14a}\\[1ex]
\text{(ii)}&~
\text{Initial condition}&
&\mathbb{K}(x, y; 0) = \delta(x - y)\mathbb{I};& \label{eq:2.14b}\\
\text{(iii)}&~
\text{Composition rule}&
&\int_{0}^{\infty}\!\!\!\mathrm{d}z\,\mathbb{K}(x,z; T_{1})\mathbb{K}(z, y; T_{2})
= 	\mathbb{K}(x, y; T_{1} + T_{2});& \label{eq:2.14c}\\
\text{(iv)}&~
\text{Unitarity}&
&\mathbb{K}^{\dagger}(x, y; T) = \mathbb{K}(y, x; -T);& \label{eq:2.14d}\\[1ex]
\text{(v)}&~
\text{Boundary condition}&
&(\mathbb{I} - \mathbb{U})\mathbb{K}(0, y; T)
- iL_{0}(\mathbb{I} + \mathbb{U})(\partial_{x}\mathbb{K})(0, y; T) = 0.& \label{eq:2.14e}
\end{alignat}
\end{subequations}
By substituting the solutions \eqref{eq:2.8} \eqref{eq:2.10} into \eqref{eq:2.13} the Feynman kernel can be written as the following momentum integral
\begin{align}
\mathbb{K}(x, y; T)
&= 	\left(
	\int_{-\infty}^{\infty}\!\frac{\mathrm{d}p}{2\pi}
	+ \oint_{C}\frac{\mathrm{d}p}{2\pi}
	\right)
	\Tilde{\mathbb{K}}(x, y; T \mid p), \label{eq:2.15}
\end{align}
where $\Tilde{\mathbb{K}}$ is the momentum representation of the kernel and given by
\begin{align}
\Tilde{\mathbb{K}}(x, y; T \mid p)
&= 	\mathbb{I}\exp\left(iT\left[p\left(\frac{x-y}{T}\right) - p^{2}\right]\right) \nonumber\\
&	+ \mathbb{S}(p)\exp\left(iT\left[p\left(\frac{x+y}{T}\right) - p^{2}\right]\right). \label{eq:2.16}
\end{align}
The integration contour $C$ is chosen to enclose all bound state poles on the complex $p$-plain in clockwise direction; see Figure \ref{fig:2}.
The contour integral picks up the bound state contributions to the Feynman kernel such that it will vanish if the S-matrix has no pole on the positive imaginary $p$-axis.

\begin{figure}[t]
\centering
\subfigure[Direct path (weight factor: $1$).]{\includegraphics{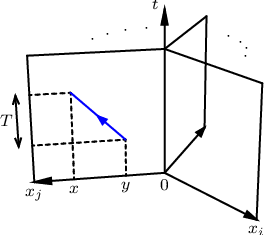} \label{fig:3a}} \quad
\subfigure[Reflected path (weight factor: $\mathbb{S}_{jj}$).]{\includegraphics{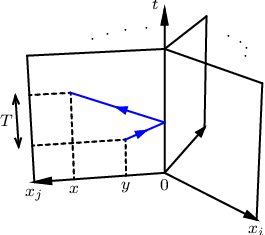} \label{fig:3b}} \quad
\subfigure[Transmitted path (weight factor: $\mathbb{S}_{ij}$).]{\includegraphics{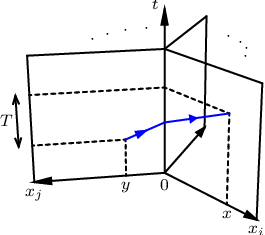} \label{fig:3c}}
\caption{Classical worldlines for a free particle on a star graph in the unfolding picture.}
\label{fig:3}
\end{figure}

\begin{figure}[t]
\centering
\subfigure[Direct path (weight factor: $\delta_{ij}$).]{\includegraphics{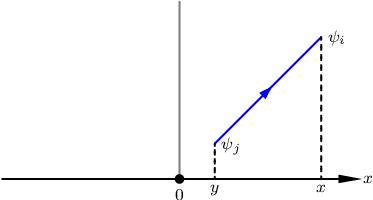} \label{fig:4a}} \quad
\subfigure[Reflected path (weight factor: $\mathbb{S}_{ij}$).]{\includegraphics{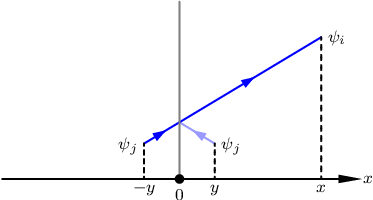} \label{fig:4b}}
\subfigure[Typical quantum fluctuation of (a) (weight factor: $\delta_{ij}$).]{\includegraphics{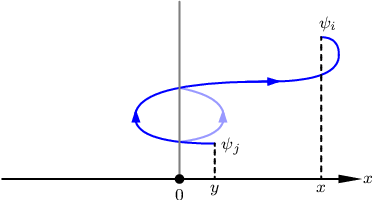} \label{fig:4c}} \quad
\subfigure[Typical quantum fluctuation of (b) (weight factor: $\mathbb{S}_{ij}$).]{\includegraphics{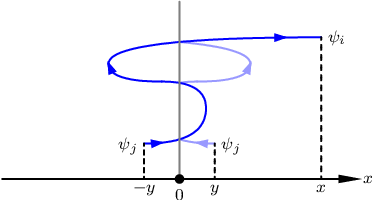} \label{fig:4d}}
\caption{Classical worldlines and its typical quantum fluctuations for a free particle on a star graph in the folding picture. Time is flowing along the vertical direction.}
\label{fig:4}
\end{figure}

The physical interpretation of \eqref{eq:2.16} is now clear: the exponent $T[p((x\mp y)/T) - p^{2}]$, which is invariant under the parity $(x, y, p) \mapsto (-x, -y, -p)$, is nothing but the classical action for a free particle propagating along the classical trajectory $x_{\text{cl}}(t) = \pm y + ((x \mp y)/T)t$ with fixed conserved momentum $p_{\text{cl}}(t) = p$, where for $x_{\text{cl}}(t) = -y + ((x + y)/T)t < 0$ it should be regarded as a mirror image $x_{\text{cl}}(t) = y - ((x + y)/T)t$ with flipped momentum $p_{\text{cl}}(t) = -p$; see Figure \ref{fig:3} and \ref{fig:4}.

By using the kernel \eqref{eq:2.15} time-evolution of any state $\Vec{\psi} \in L^{2}(\mathbb{R}^{+}) \otimes \mathbb{C}^{N}$ that satisfies the boundary condition \eqref{eq:2.5} is given by
\begin{align}
\Vec{\psi}(x, T)
&= 	\int_{0}^{\infty}\!\!\!\mathrm{d}y
	\left(
	\int_{-\infty}^{\infty}\!\frac{\mathrm{d}p}{2\pi}
	+ \oint_{C}\frac{\mathrm{d}p}{2\pi}
	\right)
	\Tilde{\mathbb{K}}(x, y; T \mid p)
	\Vec{\psi}(y). \label{eq:2.17}
\end{align}

Now we wish to represent $\Tilde{\mathbb{K}}(x, y; T \mid p)$ as a functional integral over all possible paths fluctuated around the classical trajectories.
To this end, let $t_{0} = 0 < t_{1} < \cdots < t_{n} < t_{n+1} = T$ ($t_{\ell} = \ell\delta t$ with $\delta t := T/(n+1)$) be the equidistant time sequence.
With this time sequence the exponential weighting factor $\exp(iT[p((x\mp y)/T)-p^{2}])$ can be expressed as the following time-slicing representation:
\begin{align}
& 	\exp\left(iT\left[p\left(\frac{x \mp y}{T}\right) - p^{2}\right]\right) \nonumber\\
&= 	\left[\prod_{\ell=1}^{n}\int_{-\infty}^{\infty}\!\!\!\mathrm{d}x_{\ell}\right]
	\left[\prod_{\ell=1}^{n}\int_{-\infty}^{\infty}\!\frac{\mathrm{d}p_{\ell}}{2\pi}\right]
	\exp
	\left(
	i\sum_{\ell=0}^{n}\delta t\left[p_{\ell}\left(\frac{x_{\ell+1} - x_{\ell}}{\delta t}\right) - p_{\ell}^{2}\right]
	\right)\Biggr|_{x_{0} = \pm y;~p_{0} = p}^{x_{n+1} = x}, \label{eq:2.18}
\end{align}
which is of course invariant under the parity $(x_{\ell+1}, x_{\ell}, p_{\ell}) \mapsto (-x_{\ell+1}, -x_{\ell}, -p_{\ell})$.
It should be emphasized that the integration range of $x$-coordinate is enlarged to the whole line $\mathbb{R}$ such that the path integration must be performed over paths including those depicted in Figure \ref{fig:4c} and \ref{fig:4d}.
Now, by taking the limit $n\to\infty$ (keeping $T$ fixed) we get the following phase space path integral over the path $(x, p): [0, T] \to \mathbb{R} \times \mathbb{R}$ with the boundary conditions $x(0) = \pm y$, $x(T) = x$ and $p(0) = p$:
\begin{align}
\Tilde{\mathbb{K}}(x, y; T \mid p)
&= 	\mathbb{I}
	\int_{x(0) = y}^{x(T) = x}\!\!\!\mathcal{D}x(t)
	\int_{p(0) = p}\!\!\!\mathcal{D}p(t)\,
	\exp
	\bigl(
	iS_{\text{free}}[x(t), p(t)]
	\bigr) \nonumber\\
& 	+
	\mathbb{S}(p)
	\int_{x(0) = -y}^{x(T) = x}\!\!\!\mathcal{D}x(t)
	\int_{p(0) = p}\!\!\!\mathcal{D}p(t)\,
	\exp
	\bigl(
	iS_{\text{free}}[x(t), p(t)]
	\bigr), \label{eq:2.19}
\end{align}
where $S_{\text{free}}[x(t), p(t)]$ is the action for a free particle in the Hamiltonian formulation
\begin{align}
S_{\text{free}}[x(t), p(t)]
&= 	\int_{0}^{T}\!\!\!\mathrm{d}t
	\left[p(t)\Dot{x}(t) - H_{\text{free}}\bigl(p(t)\bigr)\right], \quad
H_{\text{free}}\bigl(p(t)\bigr)
= 	p^{2}(t), \label{eq:2.20}
\end{align}
with dot $(\Dot{\phantom{x}})$ indicating the derivative with respect to $t$.
(Recall that we are working in the units $\hbar = 2m = 1$).
The integration measures are given by
\begin{subequations}
\begin{align}
\int\!\mathcal{D}x(t)
&:= 	\lim_{n\to\infty}
	\prod_{\ell=1}^{n}\left[\int_{-\infty}^{\infty}\!\!\!\mathrm{d}x_{\ell}\right], \label{eq:2.21a}\\
\int\!\mathcal{D}p(t)
&:= 	\lim_{n\to\infty}
	\prod_{\ell=1}^{n}\left[\int_{-\infty}^{\infty}\!\frac{\mathrm{d}p_{\ell}}{2\pi}\right]. \label{eq:2.21b}
\end{align}
\end{subequations}
Notice that the total numbers of $\mathrm{d}x_{\ell}$ and $\mathrm{d}p_{\ell}$ in the measures are the same as opposed to the standard phase space path integral, where $\mathcal{D}p$ contains one more $\mathrm{d}p_{\ell}$ compared to $\mathcal{D}x$.

We emphasize that, although rewriting the kernel \eqref{eq:2.13} into the path integral \eqref{eq:2.19} is an almost trivial exercise, equation \eqref{eq:2.19} is very suggestive in a sense that all the $N^{2}$ parameters which characterize the vertex are brought by the unitary weight factors $\{\mathbb{I}, \mathbb{S}(p)\}$ associated with the two distinct path classes on the half-line.
This fact strongly encourages us to understand the unitary matrices $\{\mathbb{I}, \mathbb{S}(p)\}$ as a unitary representation of a certain discrete group that all possible paths on $\mathbb{R}^{+}$ form, just like the standard weight factors in a non-simply connected space \cite{Schulman:1968yv,Laidlaw:1970ei,Schulman:1971,Dowker:1972np,Leinaas:1977fm,Horvathy:1988vh,Tanimura:1996cg,Schulman:1981}.
This naive expectation will be more transparent and turn out to be true when the S-matrix becomes independent of the momentum $p$, which is the main subject of this paper and will be discussed in detail below.

\section{Free particle at criticality} \label{sec:3}
A remarkable simplicity occurs when the S-matrix does not depend on $p$, which can be realized as fixed points of boundary RG flow.
Such boundary RG flow has first been studied in \cite{Asorey:2007rt,Asorey:2007kw} (but with different interpretations of UV and IR fixed points) and later in \cite{Ohya:2010zm}.
Search for universality classes of boundary conditions is very important not only for theoretical interests but also for physical applications to transport phenomena of Tomonaga-Luttinger (TL) liquids on quantum wire junctions.
Enormous studies have been conducted to elucidate the rich fixed point structure of TL liquid junctions (see for example \cite{Nayak:1997,Lal:2002,Oshikawa:2006,Bellazzini:2008fu}), however, there is no universally accepted phase diagram in the community thus far.
A geometric approach to scale-invariant boundary conditions which we will present in the next Section may shed new light on this subject.
We note that scale-invariant subfamily of boundary conditions have also been discussed in the literature \cite{Albeverio:1998,Fulop:1999pf,Cheon:2000tq,Fulop:2003,Cheon:2010} independently of the renormalization group study.

Now, let us move on to the analysis of path integral at criticality.
As shown in \cite{Ohya:2010zm}, boundary conditions at the fixed points are characterized by a hermitian unitary matrix satisfying the condition $\mathbb{U}^{2} = \mathbb{I}$ and parameterized by $N^{2} - N = N(N-1)$ real parameters.
There are $2^{N}$ distinct fixed points in the theory space, whose relevancy or irrelevancy are classified by the number of $+1$-eigenvalues (or $-1$-eigenvalues).
At those fixed points, the S-matrix becomes identical to $\mathbb{U}$ and the momentum integrations \eqref{eq:2.18} and \eqref{eq:2.15} are easily performed.
The resultant kernel can be written as the following configuration space path integral over the path $x: [0, T] \to \mathbb{R}$ with the boundary conditions $x(0) = \pm y$, $x(T) = x$:
\begin{align}
\mathbb{K}(x, y; T)
&= 	\mathbb{I}\int_{x(0) = y}^{x(T) = x}\!\!\!\mathcal{D}x(t)\,
	\exp
	\bigl(
	iS_{\text{free}}[x(t)]
	\bigr) \nonumber\\
&	+
	\mathbb{U}\int_{x(0) = -y}^{x(T) = x}\!\!\!\mathcal{D}x(t)\,
	\exp
	\bigl(
	iS_{\text{free}}[x(t)]
	\bigr), \label{eq:3.1}
\end{align}
where $S_{\text{free}}[x(t)]$ is the action for a free particle in the Lagrangian formulation and given by
\begin{align}
S_{\text{free}}[x(t)]
&= 	\int_{0}^{T}\!\!\!\mathrm{d}t\,\frac{1}{4}\Dot{x}^{2}(t). \label{eq:3.2}
\end{align}
The integration measure now becomes that for standard configuration space path integral (so-called Feynman measure)
\begin{align}
\int\!\mathcal{D}x(t)
&:= 	\lim_{n\to\infty}
	\frac{1}{\sqrt{4\pi i\delta t}}\prod_{\ell=1}^{n}
	\left[\int_{-\infty}^{\infty}\!\frac{\mathrm{d}x_{\ell}}{\sqrt{4\pi i\delta t}}\right], \quad
\delta t
= 	\frac{T}{n+1}. \label{eq:3.3}
\end{align}
Notice that the boundary condition for the kernel \eqref{eq:3.1} becomes two independent equations $(\mathbb{I} - \mathbb{U})\mathbb{K}(0, y; T) = 0$ and  $(\mathbb{I} + \mathbb{U})(\partial_{x}\mathbb{K})(0, y; T) = 0$, which follow from the boundary condition \eqref{eq:2.14e} and the orthonormality relations $[(\mathbb{I} \pm \mathbb{U})/2][(\mathbb{I} \mp \mathbb{U})/2] = 0$ and $[(\mathbb{I} \pm \mathbb{U})/2]^{2} = (\mathbb{I} \pm \mathbb{U})/2$ for a hermitian unitary matrix $\mathbb{U}$; see also the discussion below \eqref{eq:4.8a} and \eqref{eq:4.8b}.

Now it is easy to perform the path integration \eqref{eq:3.1}, or, equivalently, momentum integration \eqref{eq:2.15} with \eqref{eq:2.16}.
A straightforward computation gives
\begin{align}
\mathbb{K}(x, y; T)
&= 	\frac{1}{\sqrt{4\pi iT}}
	\left\{
	\mathbb{I}
	\exp\left[iT\,\frac{1}{4}\left(\frac{x - y}{T}\right)^{2}\right]
	+
	\mathbb{U}
	\exp\left[iT\,\frac{1}{4}\left(\frac{x + y}{T}\right)^{2}\right]
	\right\}. \label{eq:3.4}
\end{align}
Notice that the exponent $(T/4)((x \mp y)/T)^{2}$, which is again invariant under the parity $(x, y) \mapsto (-x, -y)$, is nothing but the classical action $S_{\text{free}}[x_{\text{cl}}(t)]$ for a free particle traveling along the classical trajectory $x_{\text{cl}}(t) = \pm y + ((x \mp y)/T)t$.
The overall factor $1/\sqrt{4\pi iT}$, on the other hand, is the standard contribution from the path integral $\int_{q(0)=q(T)=0}\mathcal{D}q(t)\exp(iS_{\text{free}}[q(t)])$, which follows from the redefinition of the path $x(t) = x_{\text{cl}}(t) + q(t)$, such that it can be regarded as ``one-loop correction'' \cite{Bastianelli:2006}.
As we will show in the next Section, $N \times N$ matrices $\{\mathbb{I}, \mathbb{U}\}$ can be derived as an $N$-dimensional unitary representation of the cyclic group $\mathbb{Z}_{2}$, whose origin is the geometry of half-line.
In this sense quantum mechanics for a free particle on a star graph at a fixed point of boundary RG flow  can be regarded as ``classical'' $+$ ``one-loop correction'' $+$ ``$\mathbb{Z}_{2}$ factor''.

We note that when $N=1$ we have only two choices for the hermitian unitary weight factor, $\mathbb{U} = +1$ or $\mathbb{U} = -1$, the former corresponds to the Neumann boundary (UV fixed points of boundary RG flow \cite{Ohya:2010zm}) and the latter the Dirichlet boundary (IR fixed points of boundary RG flow \cite{Ohya:2010zm}), which coincide with the well-known results.

\section{Adding bulk interaction: A geometric perspective} \label{sec:4}
So far we have studied path integral formulation for a free particle on a star graph by using the time-slicing representation of the kernel obtained from the operator formalism.
We have shown that all of the information about the boundary conditions can be encoded into the weight factors.
From the path integral standpoint, however, it is more desirable to formulate the weight factors without using the eigenfunctions of Hamiltonian nor the boundary conditions for wave functions.
In this Section we will show that, without referring to the boundary conditions \eqref{eq:2.5}, the momentum independent weight factors $\{\mathbb{I}, \mathbb{U}\}$ can be derived as an $N$-dimensional unitary representation of the cyclic group $\mathbb{Z}_{2} = \{I, P\}$, where $I$ (identity) and $P$ (parity) satisfy the following multiplication rules
\begin{align}
I \circ I = P \circ P = I, \quad
I \circ P = P \circ I = P. \label{eq:4.1}
\end{align}
Here circle ($\circ$) stands for abstract group multiplication.
$I$ and $P$ act on the coordinate space as usual, $I: x \mapsto x$ and $P: x \mapsto -x$.
We will see that $I$ corresponds to the unit matrix $\mathbb{I}$ and $P$ the hermitian unitary matrix $\mathbb{U}$.

\begin{figure}[t]
\centering
\subfigure[{$[\text{even}]$}: Class of loops that cross the origin even number of times.]{\quad\includegraphics{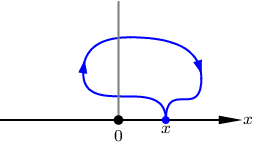}\quad \label{fig:5a}} \quad
\subfigure[{$[\text{odd}]$}: Class of loops that cross the origin odd number of times.]{\quad\includegraphics{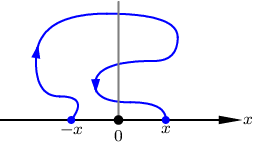}\quad \label{fig:5b}}
\subfigure[{$[\text{odd}]\ast[\text{odd}] = [\text{even}]$}: The first arrow follows from $\mathbb{Z}_{2}$ symmetry of the upper loop.]
{
\includegraphics{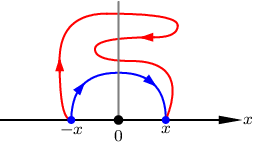}\raisebox{1.2cm}{~$\rightarrow$~}
\includegraphics{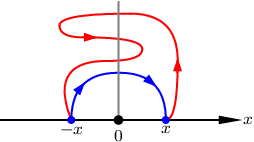}\raisebox{1.2cm}{~$\rightarrow$~}
\includegraphics{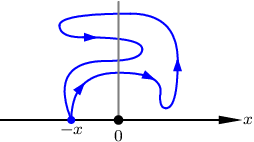} \label{fig:5c}
}
\caption{Two distinct loop classes on $\mathbb{R}/(x \sim -x)$, where the base point $x$ is chosen to exclude the origin $x=0$. (The case $x=0$ falls into the problem of boundary conditions \eqref{eq:4.8a} and \eqref{eq:4.8b}.) Composition of these loop classes satisfies the $\mathbb{Z}_{2}$ multiplication rules: $[\text{even}] \ast [\text{even}] = [\text{odd}] \ast [\text{odd}] = [\text{even}]$, $[\text{even}] \ast [\text{odd}] = [\text{odd}] \ast [\text{even}] = [\text{odd}]$, where $\ast$ indicates composition of paths.}
\label{fig:5}
\end{figure} 

To this end, it is worthwhile to remember first the standard argument of weight factors in path integral, which can be summarized as the Schulman-Laidlaw-DeWitt theorem \cite{Schulman:1968yv,Laidlaw:1970ei} (see also \cite{Schulman:1971,Dowker:1972np,Leinaas:1977fm,Horvathy:1988vh,Tanimura:1996cg,Schulman:1981}).
In a non-simply connected space $\mathcal{M}$, the total amplitude for a given transition is obtained by a linear combination of partial amplitudes, where each partial amplitude is constructed from configuration space path integral on the universal covering space of $\mathcal{M}$ over paths that belong to the same homotopy class.
The Schulman-Laidlaw-DeWitt theorem states that the coefficients in the linear combination must be a one-dimensional unitary representation of the fundamental group $\pi_{1}(\mathcal{M})$.\footnote{Horvathy \textit{et al.} discussed a generalized case where, if wave function becomes vector-valued due to some internal symmetries, weight factor could become a higher-dimensional unitary representation of $\pi_{1}(\mathcal{M})$ \cite{Horvathy:1988vh}. This is very similar to our situation although there is no genuine internal symmetry in our problem.}
Although the half-line is a simply-connected space and its fundamental group is trivial, as discussed in the previous Sections and is widely known under the name of method of images, path integral on $\mathbb{R}^{+}$ should be lifted to that on the whole line $\mathbb{R}$ with an identification $x \sim -x$ and given by a linear combination of partial amplitudes associated with the two distinct path classes on $\mathbb{R}/(x \sim -x)$, one is a class of paths that cross the origin even number of times and the other odd number of times \cite{Farhi:1989jz}.
It is intuitively clear that, for fixed initial and final positions $(x(T), x(0))$, these two path classes form the cyclic group of order 2 under the composition of paths, and are distinguished by $\mathbb{Z}_{2}$-action on the initial position $x(0) = y$ ($y>0$); see Figure \ref{fig:5} for the case of loops $(x(T), x(0)) = (x, \pm x)$.
Now that space of all possible paths we should integrate over is equipped with the group structure $\mathbb{Z}_{2}$, along the same philosophy of the Schulman-Laidlaw-DeWitt theorem, the weight factors in the linear combinations must fall into the one-dimensional unitary representation of $\mathbb{Z}_{2}$.
There are only two possibilities of such unitary representations, $\{+1, +1\}$ (trivial representation) or $\{+1, -1\}$ (defining representation), the former gives the weight factors for the Neumann boundary and the latter for the Dirichlet boundary.
Similar arguments are given in \cite{Tanimura:2001rt} by regarding the half-line as an orbifold $\mathbb{R}/\mathbb{Z}_{2}$ from the beginning.

Now we wish to extend the Schulman-Laidlaw-DeWitt theorem to the path integral on a star graph.
An important point to note is that, in the folding picture, the configuration space is $\mathbb{R}^{+}$ but the Hilbert space becomes $\mathcal{H} = L^{2}(\mathbb{R}^{+}) \otimes \mathbb{C}^{N}$ such that the Feynman kernel $\mathbb{K}$ that acts on $\mathcal{H}$ must be an $N \times N$ matrix function.
Furthermore, if the bulk interaction is edge-independent and Hamiltonian $\mathbb{H}$ is given by $\mathbb{H} = H \otimes \mathbb{I}$, which is the case of free particle, $\mathbb{H}$ and boundary conditions must be simultaneously diagonalizable such that the kernel should be factorized as $\mathbb{K} =  (\text{$N \times N$ constant matrix}) \times (\text{scalar kernel})$.
Thus we can start with the following ansatz
\begin{align}
\mathbb{K}(x, y; T)
&= 	\mathbb{W}(I)K_{\mathbb{R}}(x, y; T)
	+
	\mathbb{W}(P)K_{\mathbb{R}}(x, -y; T),
	\quad x,y > 0, \label{eq:4.2}
\end{align}
where $\mathbb{W}(I)$ and $\mathbb{W}(P)$ are $N \times N$ constant matrices to be determined later.
$K_{\mathbb{R}}(x, y; T) = \langle x|\mathrm{e}^{-iHT}|y\rangle$ is the scalar kernel defined on the whole line $\mathbb{R}$ and assumed to satisfy the following conditions:
\begin{subequations}
\begin{alignat}{3}
\text{(i)}&~
\text{Schr\"odinger equation}&
&\bigl[i\partial_{T} - H(x,-i\partial_{x})\bigr]K_{\mathbb{R}}(x, y; T) = 0;& \label{eq:4.3a}\\[1ex]
\text{(ii)}&~
\text{Initial condition}&
&K_{\mathbb{R}}(x, y; 0) = \delta(x - y);& \label{eq:4.3b}\\
\text{(iii)}&~
\text{Composition rule}&\quad
&\int_{-\infty}^{\infty}\!\!\!\mathrm{d}z\,K_{\mathbb{R}}(x,z; T_{1})K_{\mathbb{R}}(z, y; T_{2})
= 	K_{\mathbb{R}}(x, y; T_{1} + T_{2});& \label{eq:4.3c}\\
\text{(iv)}&~
\text{Unitarity}&
&K_{\mathbb{R}}^{\ast}(x, y; T) = K_{\mathbb{R}}(y, x; -T);& \label{eq:4.3d}\\[1ex]
\text{(v)}&~
\text{$\mathbb{Z}_{2}$ symmetry}&
&K_{\mathbb{R}}(x, y; T) = K_{\mathbb{R}}(-x, -y; T),& \label{eq:4.3e}
\end{alignat}
\end{subequations}
where asterisk (${}^{\ast}$) indicates complex conjugate.
Equation \eqref{eq:4.3a} assures the total amplitude \eqref{eq:4.2} to satisfy the Schr\"odinger equation $[i\partial_{T}\mathbb{I} - \mathbb{H}(x, -i\partial_{x})]\mathbb{K}(x, y; T) = 0$.
We note that $\mathbb{Z}_{2}$ symmetry \eqref{eq:4.3e} will be justified for parity invariant bulk Hamiltonian $H$.
Construction of scalar kernel $K_{\mathbb{R}}$ satisfying the conditions \eqref{eq:4.3a}--\eqref{eq:4.3e} is a standard problem so that we assume $K_{\mathbb{R}}$ is given by an ordinary configuration space path integral.

With these assumptions, the total amplitude $\mathbb{K}$ will satisfy the initial condition \eqref{eq:2.14b}, composition rule \eqref{eq:2.14c} and unitarity \eqref{eq:2.14d} if and only if the weight factors form an $N$-dimensional unitary representation of the cyclic group $\mathbb{Z}_{2}$.
Let us prove this statement by imposing these conditions to the kernel $\mathbb{K}$.
\begin{itemize}
\item
\textit{Initial condition.}
Let us first impose the initial condition \eqref{eq:2.14b} to $\mathbb{K}$.
A straightforward calculation gives $\mathbb{K}(x, y; 0) = \mathbb{W}(I)\delta(x-y) + \mathbb{W}(P)\delta(x+y)$, where we have used the assumption \eqref{eq:4.3b}.
Since $\delta(x+y) = 0$ for $x,y>0$, the initial condition \eqref{eq:2.14b} leads to the following constraint
\begin{align}
&\mathbb{W}(I) = \mathbb{I}. \label{eq:4.4}
\end{align}

\item
\textit{Composition rule.}
Let us next impose the composition rule \eqref{eq:2.14c} to $\mathbb{K}$.
By substituting the ansatz the left hand side of \eqref{eq:2.14c} becomes $\int_{0}^{\infty}\!\!\mathrm{d}z\,\mathbb{K}(x,z;T_{1})\mathbb{K}(z,y;T_{2}) = \bigl(\mathbb{W}(I)\mathbb{W}(I)\int_{0}^{\infty}\!\!\mathrm{d}z + \mathbb{W}(P)\mathbb{W}(P)\int_{-\infty}^{0}\!\mathrm{d}z\bigr)K_\mathbb{R}(x,z;T_{1})K_{\mathbb{R}}(z,y;T_{2}) + \bigl(\mathbb{W}(I)\mathbb{W}(P)\int_{0}^{\infty}\!\!\mathrm{d}z + \mathbb{W}(P)\mathbb{W}(I)\int_{-\infty}^{0}\!\mathrm{d}z\bigr)K_{\mathbb{R}}(x,z;T_{1})K_{\mathbb{R}}(z,-y;T_{2})$, where we have used \eqref{eq:4.3e}.
In order to realize the right hand side of \eqref{eq:2.14c} the weight factors must obey the following multiplication rules
\begin{subequations}
\begin{align}
&\mathbb{W}(I)\mathbb{W}(I) = \mathbb{W}(P)\mathbb{W}(P) = \mathbb{W}(I), \label{eq:4.5a}\\
&\mathbb{W}(I)\mathbb{W}(P) = \mathbb{W}(P)\mathbb{W}(I) = \mathbb{W}(P), \label{eq:4.5b}
\end{align}
\end{subequations}
which are nothing but those for $\mathbb{Z}_{2}$; see \eqref{eq:4.1}.
Thus the weight factors $\{\mathbb{W}(I), \mathbb{W}(P)\}$ must be an $N \times N$ matrix representation of the cyclic group $\mathbb{Z}_{2}$.

\item
\textit{Unitarity.}
Let us finally impose the unitarity condition \eqref{eq:2.14d} to $\mathbb{K}$.
By taking the hermitian conjugate of the ansatz we get $\mathbb{K}^{\dagger}(x,y;T) = \mathbb{W}^{\dagger}(I)K_{\mathbb{R}}(y, x; -T) + \mathbb{W}^{\dagger}(P)K_{\mathbb{R}}(y, -x; -T)$, where we have used \eqref{eq:4.3d} and \eqref{eq:4.3e}.
Thus the unitarity leads to the following constraints
\begin{align}
&\mathbb{W}^{\dagger}(I) = \mathbb{W}(I), \quad
\mathbb{W}^{\dagger}(P) = \mathbb{W}(P). \label{eq:4.6}
\end{align}
Hence the weight factors $\{\mathbb{W}(I), \mathbb{W}(P)\}$ must be hermitian matrices.
It follows immediately from \eqref{eq:4.4}, \eqref{eq:4.5a} and \eqref{eq:4.6} that the weight factors must also be unitary, $\mathbb{W}^{-1}(I) = \mathbb{W}^{\dagger}(I)$ and $\mathbb{W}^{-1}(P) = \mathbb{W}^{\dagger}(P)$.
\end{itemize}
Collecting the above results we conclude that $\mathbb{W}$ must be an $N$-dimensional unitary representation of the cyclic group $\mathbb{Z}_{2}$; that is, it must be a map $\mathbb{W}: \mathbb{Z}_{2} \to U(N)$.
Such unitary representation is given by
\begin{align}
\mathbb{W}(I) = \mathbb{I}, \quad
\mathbb{W}(P) = \mathbb{U}, \label{eq:4.7}
\end{align}
where $\mathbb{U} \in U(N)$ is a hermitian unitary matrix that satisfies $\mathbb{U}^{2} = \mathbb{I}$.
Different choice of $\mathbb{U}$ corresponds to different quantization.

We emphasize that the weight factors are determined without referring to the boundary conditions.
This is very important because the path integral presented in the previous Section is based on the boundary conditions provided by the self-adjoint extension of \textit{free} Hamiltonian, while the results \eqref{eq:4.7} do not depend on any specific Hamiltonian (except for the parity invariance).
Indeed, boundary conditions for $\mathbb{K}$ can be derived as a consequence of \eqref{eq:4.7}:
If the scalar kernel is continuous and smooth at the origin, $K_{\mathbb{R}}(0_{+}, y;T) = K_{\mathbb{R}}(0_{-}, y;T)$ and $(\partial_{x}K_{\mathbb{R}})(0_{+}, y;T) = (\partial_{x}K_{\mathbb{R}})(0_{-}, y;T)$, which will be justified when there is no singularity at $x=0$ in the parity invariant bulk Hamiltonian, we immediately obtain the following boundary conditions for the total Feynman kernel:
\begin{subequations}
\begin{align}
& 	(\mathbb{I} - \mathbb{U})\mathbb{K}(0_{+}, y;T)
= 	(\mathbb{I} - \mathbb{U})(\mathbb{I} + \mathbb{U})K_{\mathbb{R}}(0_{+}, y;T)
= 	0, \label{eq:4.8a}\\
& 	(\mathbb{I} + \mathbb{U})(\partial_{x}\mathbb{K})(0_{+}, y;T)
= 	(\mathbb{I} + \mathbb{U})(\mathbb{I} - \mathbb{U})(\partial_{x}K_{\mathbb{R}})(0_{+}, y;T)
= 	0, \label{eq:4.8b}
\end{align}
\end{subequations}
where we have used $\mathbb{Z}_{2}$ symmetry $K_{\mathbb{R}}(x, -y; T) = K_{\mathbb{R}}(-x, y; T)$.
Equations \eqref{eq:4.8a} and \eqref{eq:4.8b} are precisely the same boundary conditions for a free particle at the fixed point of boundary RG flow.

To summarize, we have shown that one-particle quantum mechanics on a star graph with edge-independent bulk interaction at a fixed point of boundary RG flow is formulated into the following configuration space path integral
\begin{align}
\mathbb{K}(x, y; T)
&= 	\sum_{n=0,1}\mathbb{U}^{n}
	\int_{x(0) = (-1)^{n}y}^{x(T) = x}\!\!\!\mathcal{D}x(t)\,
	\exp
	\bigl(
	iS[x(t)]
	\bigr),
	\quad \mathbb{U} \in U(N),
	\quad \mathbb{U}^{2} = \mathbb{I}, \label{eq:4.9}
\end{align}
where $S[x(t)]$ is a generic parity invariant one-particle action in the Lagrangian formulation and given by $S[x(t)] = \int_{0}^{T}\!\mathrm{d}t\,L(x(t), \Dot{x}(t))$.
The most important Lagrangians for physical applications will be as follows:
\begin{align}
L\bigl(x(t), \Dot{x}(t)\bigr)
&= 	\begin{cases}
	\displaystyle
	\frac{1}{4}\Dot{x}^{2}(t) - V\bigl(x(t)\bigr), \\[1em]
	\displaystyle
	\frac{1}{4}\Dot{x}^{2}(t) + eA\bigl(x(t)\bigr)\Dot{x}(t),
	\end{cases} \label{eq:4.10}
\end{align}
where $V$ is a parity invariant external potential, $V(-x) = V(x)$, and $A$ is an external vector potential that transforms under the parity as $A \stackrel{P}{\mapsto} -A$, with $e$ being a gauge coupling constant.
Examples of such external potentials are harmonic potential $V(x) = (1/4)\omega^{2}x^{2}$, P\"oschl-Teller potential of $\cosh$ type $V(x) = -g/\cosh^{2}(x)$, and so on.
(Note that in this argument bulk theory does not necessarily lie on a fixed point of RG flow.)
Scalar kernel $K_{\mathbb{R}}$ for these parity invariant regular potentials can be continuous and smooth at the origin such that the boundary conditions  for the total Feynman kernel $\mathbb{K}$ is the same as \eqref{eq:4.8a} and \eqref{eq:4.8b}.
On the other hand, if one consider singular potentials such as Coulomb potential $V(x) = Ze^{2}/|x|$, P\"oschl-Teller potential of $\sinh$ type $V(x) = g/\sinh^{2}(x)$ or inverse square potential $V(x) = g/x^{2}$ (namely, conformal mechanics on star graph), the boundary conditions for $\mathbb{K}$ must be modified.
Nevertheless, the weight factors are still given by an $N$-dimensional unitary representation of $\mathbb{Z}_{2}$ as long as scalar kernel itself and its boundary conditions at the origin are both invariant under the parity $P$.

\vskip 1em
\noindent
\textbf{Time-reversal symmetry}.
Before closing this Section let us briefly discuss how symmetries constrain the weight factors $\{\mathbb{W}(I), \mathbb{W}(P)\}$.
For the sake of simplicity, we will restrict ourselves to the symmetry under the time-reversal $\mathcal{T}: T \mapsto -T$, which acts on the bulk wave function as $\mathcal{T}: \Vec{\psi}(x, T) \mapsto \Vec{\psi}^{\ast}(x, -T)$.
As is well-known \cite{Cheon:2000tq}, time-reversal symmetry can be broken by spatial boundary even if the bulk theory is invariant under $\mathcal{T}$.
We would like to classify $\mathcal{T}$-invariant subfamily of weight factors from the viewpoint of representation theory of $\mathbb{Z}_{2}$.

Since the time evolution is given by $\Vec{\psi}(x, T) = \int_{0}^{\infty}\!\mathrm{d}y\,\mathbb{K}(x, y; T)\Vec{\psi}(y)$, the system becomes $\mathcal{T}$-invariant if and only if the following condition holds
\begin{align}
\mathbb{K}^{\ast}(x, y; -T)
&= 	\mathbb{K}(x, y; T). \label{eq:4.11}
\end{align}
Notice that for a unitary kernel the condition \eqref{eq:4.11} is equivalent to $\mathbb{K}^{T}(x, y; T) = \mathbb{K}(y, x; T)$.
Hence, if the bulk scalar kernel is invariant under $\mathcal{T}$, $K_{\mathbb{R}}^{\ast}(x, y; -T) = K_{\mathbb{R}}(x, y; T)$, the condition \eqref{eq:4.11} to the ansatz \eqref{eq:4.2} leads to the following additional constraints for the weight factors
\begin{align}
\mathbb{W}^{\ast}(I)
&= 	\mathbb{W}(I), \quad
\mathbb{W}^{\ast}(P)
= 	\mathbb{W}(P), \label{eq:4.12}
\end{align}
which are equivalent to $\mathbb{W}^{T}(I) = \mathbb{W}(I) = \mathbb{W}^{-1}(I)$ and $\mathbb{W}^{T}(P) = \mathbb{W}(P) = \mathbb{W}^{-1}(P)$ for hermitian unitary matrices.
Thus, for time-reversal invariant systems, the weight factors $\{\mathbb{W}(I), \mathbb{W}(P)\}$ must be an $N$-dimensional orthogonal matrix representation of the cyclic group $\mathbb{Z}_{2}$; that is, it must be a map $\mathbb{W}: \mathbb{Z}_{2} \to O(N)$.
Such matrix representation is given by
\begin{align}
\mathbb{W}(I) = \mathbb{I}, \quad
\mathbb{W}(P) = \mathbb{O}, \label{eq:4.13}
\end{align}
where $\mathbb{O} \in O(N)$ is a symmetric orthogonal matrix that satisfies $\mathbb{O}^{2} = \mathbb{I}$.
Notice that \eqref{eq:4.13} coincides with the previous results discussed in the case $N=2$ \cite{Cheon:2000tq}.
Other symmetry constraints such as invariance under permutation of different edges can be studied analogously.

\section{Conclusions and discussions} \label{sec:5}
In this paper we studied path integral description of one-particle quantum mechanics on a star graph and proposed a weight factor formulation of boundary conditions.
Our proposal is based on the folding trick, where in the folded theory wave functions become $N$-component vector-valued functions.
In Section \ref{sec:2} we carefully studied free particle case and showed that $U(N)$ family of boundary conditions can be encoded into the momentum dependent weight factors $\{\mathbb{I}, \mathbb{S}(p)\}$ in the phase space path integral, where $\mathbb{S}(p) \in U(N)$ is a hermitian analytic unitary S-matrix that satisfies $\mathbb{S}(p)\mathbb{S}(-p) = \mathbb{I}$.
For the price of weight factor formulation, however, we have to include non-standard contour integral with respect to the initial momentum $p(0) = p$ in order to pick up the bound state contributions.
In the subsequent Section we showed that, when the theory lies on a fixed point of boundary RG flow, the path integral can be simplified to the configuration space path integral with the momentum independent weight factors $\{\mathbb{I}, \mathbb{U}\}$, where $\mathbb{U} \in U(N)$ is a hermitian unitary matrix that satisfies $\mathbb{U}^{2} = \mathbb{I}$.
In Section \ref{sec:4} we generalized to the interacting case by considering edge-independent parity invariant bulk Hamiltonian and showed that, without referring to the boundary conditions, the momentum independent weight factors are generally given by an $N$-dimensional unitary representation of the cyclic group $\mathbb{Z}_{2}$; that is, scale-invariant boundary conditions have emerged only through the $\mathbb{Z}_{2}$ structure of bulk geometry.
If the system is time-reversal invariant, the weight factors are reduced to an $N$-dimensional orthogonal matrix representation of $\mathbb{Z}_{2}$.

We believe that our results presented in this paper give a major step toward the understanding of path integral formulation on quantum graph.
We also believe that our results provide the foundation for worldline formalism \cite{Schubert:2001he} on star graphs, which is the first-quantization approach to perturbative quantum field theory.
Indeed, by using the results \eqref{eq:4.10} with \eqref{eq:4.11} it is straightforward to apply the techniques developed in \cite{Bastianelli:2008vh}.
For example, the ground state energy (Casimir energy) per unit volume at the position $(x, j)$ for relativistic $\mathcal{T}$-invariant free massless real scalar field theory on $\mathbb{R}^{D} \times (\text{star graph})$ at a fixed point of boundary RG flow is evaluated by the following Wick-rotated ($iT \to T$) configuration space path integral
\begin{align}
\mathcal{E}_{\text{Casimir}}(x, j)
&= 	-\frac{1}{2}\mathbb{O}_{jj}
	\int_{0}^{\infty}\!\frac{\mathrm{d}T}{T}\frac{1}{(4\pi T)^{D/2}}
	\int_{x(0) = -x}^{x(T) = x}\!\!\!\mathcal{D}x(\tau)\,
	\exp\left(-\int_{0}^{T}\!\!\!\mathrm{d}\tau\,\frac{1}{4}\Dot{x}^{2}(\tau)\right) \nonumber\\
&= 	-\frac{1}{2}\frac{\Gamma(\tfrac{D+1}{2})}{(4\pi x^{2})^{(D+1)/2}}\mathbb{O}_{jj},\label{eq:5.1}
\end{align}
where we have subtracted $n=0$ term, which is the contribution from the whole line, and the factor $1/(4\pi T)^{D/2}$ is the one-loop contribution on $\mathbb{R}^{D}$.
Equation \eqref{eq:5.1} coincides with the previous result \cite{Bellazzini:2006jb} obtained in the operator formalism when $D=1$.
Similarly, one can compute $n$-point Green's function, one-loop effective action and Schwinger effect for relativistic scalar field theory (with edge-independent background gauge field) on a star graph.

We are left with a number of questions, however.
One of the main questions is the following: Is it possible to derive the weight factors $\{\mathbb{I}, \mathbb{S}(p)\}$ as the momentum dependent $N$-dimensional unitary representation of the cyclic group $\mathbb{Z}_{2}$?
One may expect that, by allowing the weight factors to depend on the initial momentum $p(0) = p$ (that is, weight factors in the phase space), the initial condition, composition rule and unitarity for the momentum representation of the total Feynman kernel $\Tilde{\mathbb{K}}(x, y; T \mid p)$ will lead to the unique solution to the hermitian analytic unitary S-matrix $\mathbb{S}(p) \in U(N)$ that satisfies $\mathbb{S}(p)\mathbb{S}(-p) = \mathbb{I}$.
However, the S-matrix obviously depends on the explicit form of the bulk Hamiltonian $H$ such that in order to determine $\mathbb{S}(p)$ we need an additional information about bulk interaction.
So how should we include such information without solving the bulk Schr\"odinger equation nor using the boundary conditions?
One of candidates is to impose an additional condition that the bulk system is semiclassical and one-loop exact (or WKB exact), which is the case of free particle discussed in Section \ref{sec:2}.
Another important issue is that we have to determine the integration contour $C$ from some algebraic relations.
In our approach, determination of the S-matrix singularity structure is mandatory.
Resolutions to these problems are welcome.

\section*{Acknowledgment}
The author would like to thank Mihail Mintchev for stimulating discussions and reading the manuscript carefully.
He is also grateful to Toshiaki Fujimori, Makoto Sakamoto and Izumi Tsutsui for enlightening conversations.
Part of this work was carried out under JSPS Research Fellowships for Young Scientists and JSPS Excellent Young Researchers Overseas Visit Program.

\appendix
\section{Proof of orthonormality and completeness} \label{appendix:A}
In this Appendix  we discuss computational details for the proof of orthonormality and completeness relations of the energy eigenfunctions \eqref{eq:2.8} and \eqref{eq:2.10}.
Since the relations \eqref{eq:2.11b} and \eqref{eq:2.11c} are easily derived, we will concentrate ourselves to the proof of \eqref{eq:2.11a} and \eqref{eq:2.12}.
We note that essentially the same computations presented in this Appendix can also be found in \cite{Cheon:2000tq,Bellazzini:2006jb}.

Let us first show the orthonormality relation \eqref{eq:2.11a}.
Substituting the solution \eqref{eq:2.8} into the left hand side of \eqref{eq:2.11a} we get
\begin{align}
\int_{0}^{\infty}\!\!\!\mathrm{d}x\,
\bigl[\Vec{\psi}^{a}(x; p)\bigr]^{\dagger} \cdot \Vec{\psi}^{b}(x; q)
&= 	\delta^{ab}
	\biggl[
	2\pi\delta(p - q)
	+
	2\pi\delta(p + q)\frac{ipL^{a} + 1}{ipL^{a} - 1} \nonumber\\
&	+
	\mathcal{P}\frac{i}{p - q}
	\left(1 - \frac{ipL^{a} + 1}{ipL^{a} - 1}\frac{iqL^{a} - 1}{iqL^{a} + 1}\right) \nonumber\\
&	-
	\mathcal{P}\frac{i}{p + q}
	\left(\frac{ipL^{a} + 1}{ipL^{a} - 1} - \frac{iqL^{a} - 1}{iqL^{a} + 1}\right)
	\biggr], \label{eq:A.1}
\end{align}
where we have used the relation $(\Vec{\xi}^{a})^{\dagger} \cdot \Vec{\xi}^{b} = \delta^{ab}$ and the integration formula
\begin{align}
\int_{0}^{\infty}\!\!\!\mathrm{d}x\,
\mathrm{e}^{\pm ipx}
&= 	\pm\lim_{\epsilon \to 0_{+}}\frac{i}{p \pm i\epsilon}
= 	\pm\mathcal{P}\frac{i}{p} + \pi\delta(p). \label{eq:A.2}
\end{align}
Here $\mathcal{P}$ stands for the principal value.
It is easy to check that the last two lines in \eqref{eq:A.1} cancel each other.
Noting that $\delta(p + q) = 0$ for $p, q > 0$, we obtain the equation \eqref{eq:2.11a}.

Let us next show the completeness relation \eqref{eq:2.12}.
A straightforward calculation gives
\begin{align}
\sum_{a=1}^{N}\int_{0}^{\infty}\!\frac{\mathrm{d}p}{2\pi}\,
\Vec{\psi}^{a}(x; p) \cdot \bigl[\Vec{\psi}^{a}(y; p)\bigr]^{\dagger}
&= 	\mathbb{I}
	\int_{-\infty}^{\infty}\frac{\mathrm{d}p}{2\pi}\,
	\mathrm{e}^{ip(x - y)}
	+
	\int_{-\infty}^{\infty}\frac{\mathrm{d}p}{2\pi}\,
	\mathbb{S}(p)
	\mathrm{e}^{ip(x + y)} \nonumber\\
&= 	\delta(x - y)\mathbb{I}
	-
	\sum_{0<L^{a}<\infty}
	\frac{2}{L^{a}}\mathrm{e}^{-(x+y)/L^{a}}\mathbb{P}^{a}, \label{eq:A.3}
\end{align}
where $x,y>0$.
In the first line we have used $\Vec{\xi}^{a}(\Vec{\xi}^{a})^{\dagger} = \mathbb{P}^{a}$, $\sum_{a=1}^{N}\mathbb{P}^{a} = \mathbb{I}$ and the definition \eqref{eq:2.9}, and in the second line the contour integral with the contour along the real $p$-axis enclosing the upper half complex $p$-plain.
Since the second term of \eqref{eq:A.3} is nothing but the opposite sign of the contribution from the bound states, $\sum_{0<L^{a}<\infty}\Vec{\psi}_{B}^{a}(x) \cdot [\Vec{\psi}_{B}^{a}(y)]^{\dagger}$, we arrive at the completeness relation \eqref{eq:2.12}.
Now it is obvious that, as noted in \cite{Bellazzini:2006jb}, the condition for the absence of bound states is simply expressed as follows:
\begin{align}
\int_{-\infty}^{\infty}\frac{\mathrm{d}p}{2\pi}\,
\mathbb{S}(p)\mathrm{e}^{ipx}
= 	0
\quad\text{for}\quad
x>0. \label{eq:A.4}
\end{align}

\bibliographystyle{utphys}
\bibliography{bibliography}

\end{document}